\documentclass[twocolumn,showpacs,preprintnumbers,prl]{revtex4-1}

\usepackage{graphicx}
\usepackage{dcolumn}
\usepackage{bm}
\usepackage{amsmath}
\usepackage{amssymb}
\usepackage{hyperref}

\begin{document}
\title{Hydrodynamic long-time tails after a quantum quench}
\author{Jonathan Lux$^{1}$}
\email{lux@thp.uni-koeln.de}
\author{Jan M\"uller$^1$}
\author{Aditi Mitra$^{1,2}$}
\author{Achim Rosch$^{1}$}
\affiliation{$^1$ Institut f\"ur Theoretische Physik, Universit\"at zu K\"oln, D-50937 Cologne, Germany\\
$^2$ Department of Physics, New York University, 4 Washington Place, New York, NY 10003, USA}
\date{\today}

\begin{abstract}
After a quantum quench, a sudden change of parameters, generic many particle quantum systems are expected to equilibrate. A few collisions of quasiparticles are usually sufficient to establish approximately local equilibrium. Reaching global equilibrium
 is, however, much more difficult as conserved quantities have to be
transported for long distances to build up a pattern of fluctuations characteristic for equilibrium. Here we investigate the quantum quench of the one-dimensional bosonic Hubbard model from infinite to finite interaction strength $U$ using semiclassical methods for weak, and exact diagonalization for strong quenches. Equilibrium is approached only slowly, as $t^{-1/2}$ with subleading corrections proportional to $t^{-3/4}$, consistent with predictions from hydrodynamics. We show that these long-time tails determine the relaxation of a
wide range of physical observables.
\end{abstract}

\maketitle
States at thermal equilibrium can be described with only a few macroscopic parameters like temperature, $T$, and chemical potential, $\mu$. The fundamental question of how such an equilibrium state can be reached for a interacting quantum system has recently gained a lot of
attention~\cite{Calabrese06,Kollath07,Rigol08,Kehrein08,Review11,MitraGiamarchi11,Kuhr12,Trotzky12}, partially due to new experimental opportunities to study this question using ultracold atoms. They allow to realize simple model Hamiltonians and to change their parameters practically instantaneously to study thermalization
in a closed quantum system with unprecedented precision and control.

 In a typical quantum quench experiment, one considers the evolution of the ground-state wave function, $| \Psi_0\rangle$, of an initial Hamiltonian, $H_0$, after a sudden change of the Hamiltonian, $H_0 \to H$.
The time evolution, $|\Psi(t)\rangle=e^{-i H t} |\Psi_0\rangle$, of a  many-particle system  occurs generically in three main steps: prethermalization, local equilibration, and global equilibration. First the wave function starts to adjust to the new Hamiltonian~\cite{Kehrein08, Berges04} on a short
time scale.  After this, often a quasi-stationary `prethermalized' state is obtained~\cite{Gring12,Kollar11,Mitra13,Marcuzzi13}, where quasiparticles have been formed but not yet scattered with each other. Sometimes coherent oscillations characterize this regime for large quenches~\cite{Barankov04,Yuzbashyan06,Eckstein09}. In a second step, a few scattering processes of the excitation are often sufficient to achieve approximately a local equilibrium state. This can, for example, be captured within a kinetic equation approach~\cite{SpohnKE13,Tavora13,Kollar13}. Finally the third step, the buildup of a global equilibrium after a quantum quench, has received probably the least attention and will be the focus of this paper. It is dominated by the diffusive transport of conserved quantities, like matter or energy, over large distances. It therefore leads to pronounced  long-time tails, well known from
hydrodynamics.

\begin{figure}[t]
\includegraphics[width=0.75 \linewidth]{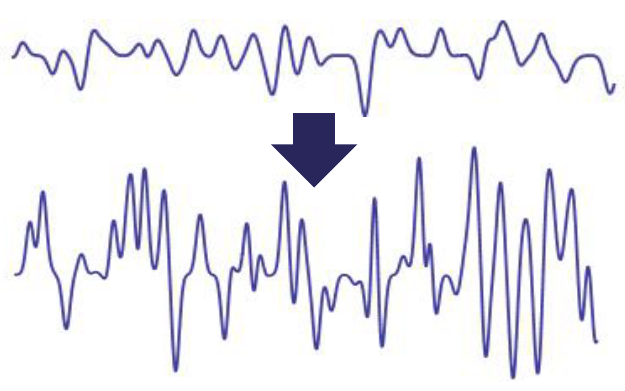}
\caption{After a quench and subsequent thermalization, the amplitude of local fluctuations of for example the energy density, changes. As energy has to be transported diffusively over large distances to build up the new pattern of fluctuations, equilibration takes long times resulting in hydrodynamic long-time tails. \label{fig1} }
\end{figure}
Two situations have to be distinguished when studying hydrodynamic long-time tails after a quantum quench: homogeneous and inhomogeneous systems. For example, if the relaxation of ultracold atoms in a trap is considered, in general the spatial distribution of, e.g., the energy will be different before and after the quench. This implies that energy has to be transported over large distances of the order of the size $L$ of the system \cite{Rapp10,kollathInhomo}. If this transport is diffusive, this takes a very long time,
$L^2/D$, where $D$ is a diffusion constant. In Ref.~\cite{Rapp10} this effect was studied quantitatively for weakly interacting fermions released from a trap.
Perhaps more surprisingly, conservation laws and diffusion  lead also to
extremely slow relaxation in translationally invariant systems studied in the following.
Here it is important to realize that any classical or quantum state is characterized by a pattern of
fluctuations. Consider, for example, fluctuations of the energy density,  $\delta e=e-\langle e\rangle$,
in a system at $T>0$, where correlations decay on a finite length scale $l_c$.  In  equilibrium and  on length scales large compared to $l_c$, they  can  be described by the equal-time correlation function
\begin{eqnarray}
\langle \delta e(\bm r) \delta e(\bm r') \rangle_{\rm eq} \approx c_v\, k_B T^2\,  \delta(\bm r - \bm r')
\end{eqnarray}
 where $c_v=(\langle H^2 \rangle-\langle H \rangle^2)/(V k_B T^2)$ is the specific heat per volume.
During equilibration, the system has to build up this fluctuation pattern, see Fig.~\ref{fig1}.
If only energy is conserved, one can describe the equilibration at long times
 by a stochastic  linearized diffusion equation
 \begin{equation}\partial_t e-D \nabla^2 e=\nabla f \label{diffusion}\end{equation}
where  $\langle f(\bm r,t) f(\bm r',t') \rangle_{\rm eq}= 2 D c_v T^2 \delta(\bm r-\bm r') \delta(t-t')$.
As discussed in the supplementary material \cite{supplement}, one obtains from a straightforward solution of this equation in
$d$ spatial dimensions
\begin{equation}\label{tails}
\langle \delta e(\bm r,t) \delta e(\bm r',t) \rangle -\langle \delta e(\bm r) \delta e(\bm r')\rangle_{\rm eq} \sim t^{-\frac{d}{2}}+ \mathcal O\!\left(t^{-\frac{3 d}{4}}\!\right)
\end{equation}
for $|\vec r - \vec r'|^2 \ll D t$.
A simple scaling analysis using Eq. (\ref{diffusion}) as a fixed point, see supplement \cite{supplement},
shows that correction terms \cite{Huse06} to Eq. (\ref{diffusion}), like $ \partial_x (e \partial_x e) $,
lead to corrections vanishing with $1/t^{3 d/4}$. The same results are obtained when additional diffusive modes (e.g., particle density $n$)  are included.
The situation is, however, different in systems with momentum conservation. In this case the momentum current (i.e., the pressure) has contributions proportional to $(\delta e)^2$ and $(\delta n)^2$.
These are relevant perturbation in dimensions $d<2$, described by the KPZ universality class \cite{KPZ,Praehofer04,Beijeren12,Spohn13} in $d=1$. In this case one expects  that some modes relax with $1/t^{2/3}$ instead of $1/t^{1/2}$ as observed numerically~\cite{Grassb02}.

In this paper we will study a quantum quench in a lattice model without momentum conservation.
We consider the $1d$ bosonic Hubbard model
\begin{eqnarray} \label{bhmodel}
H=-J \sum_i a^\dagger_i a_{i+1}+h.c. + \frac{U}{2} \sum_{i} n_i (n_i-1)
\end{eqnarray}
after a sudden quench from an initial state at $U=\infty$ where $n_i =a_i^\dagger a_i=1$
to a state with finite $U$.

We have chosen this model for four reasons: (i) The bosonic Hubbard model
is probably the many-particle model best suited for experimental quench studies using ultracold atoms~\cite{Greiner02,Kuhr12,Trotzky12}, (ii) long-time tails are most pronounced in $1d$, (iii) the $1d$ case is especially suited for numerical studies, and, finally, (iv) in contrast to many other simple $1d$ models, the bosonic Hubbard model is {\em not} close to an integrable point: the dominant excitations of the bosonic Mott insulator, doublons and holons, have a {\em different} dispersion and can  therefore equilibrate by simple two-particle collisions~\cite{Garst01}.

We first consider a weak quench from $U=\infty$ to a finite but large $U \gg J$. In this limit a dilute
gas of quasiparticles, holons (empty sites) with dispersion $\epsilon^h_k\approx -2 J \cos k$ and doublons
(doubly occupied sites) with energy $\epsilon^d_k\approx -4 J \cos k$ are created. As their average distance $\rho^{-1} = 1/4 \, (U/J)^2$ \cite{kollath12} is much larger than their typical wavelength, a simple quasi-classical treatment of their dynamics is possible
following Sachdev and Damle~\cite{Sachdev97}. This approach was recently applied to quantum quenches
in an integrable system in Ref.~[\onlinecite{Rieger11}] and to short time dynamics in Ref.~[\onlinecite{kollath12}]. While the motion  of the quasi-particles can be treated classically, their creation and scattering is a quantum mechanical process.

 \begin{figure}[t]
 \includegraphics[width=0.9 \linewidth]{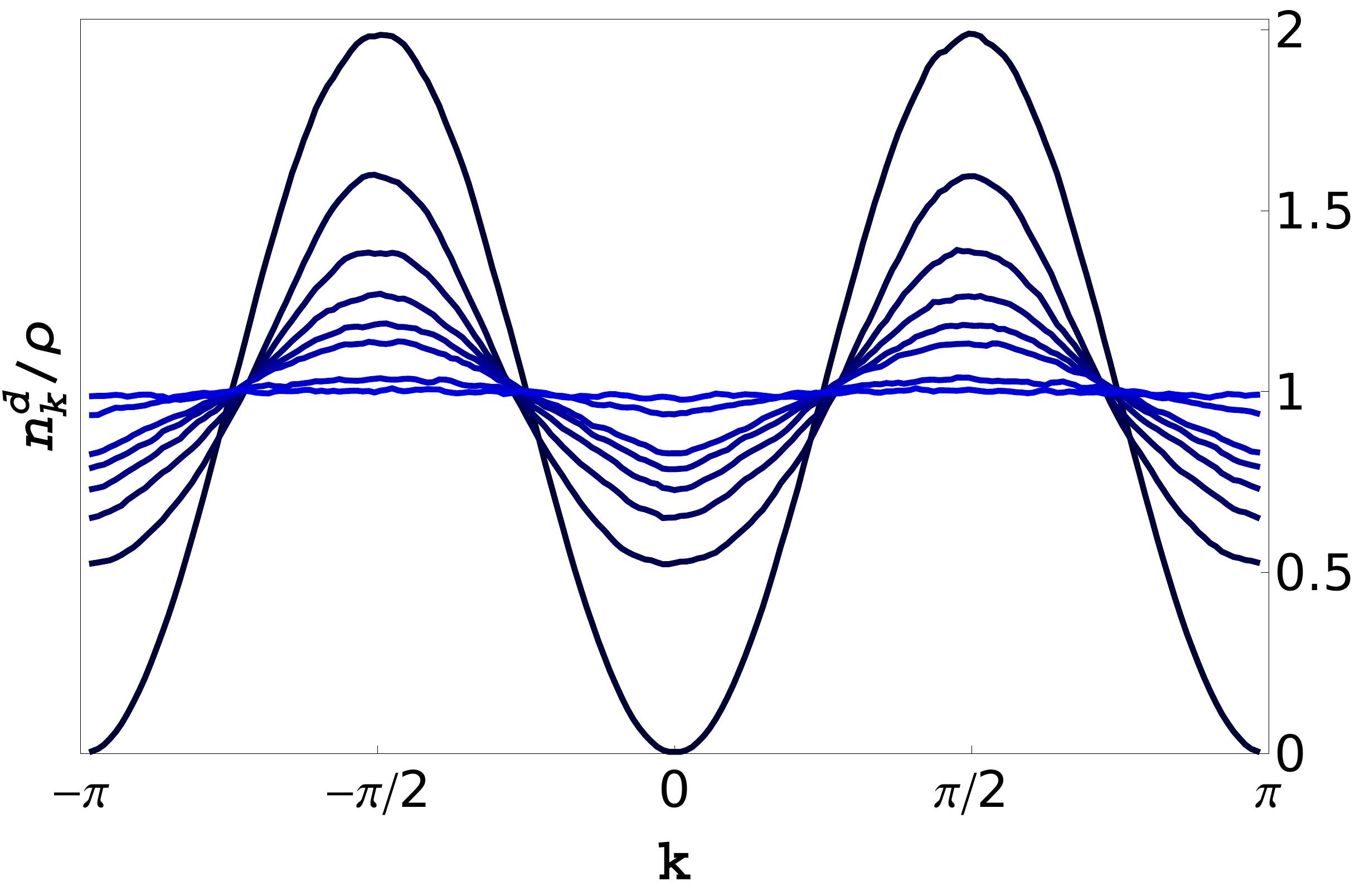}
 \caption{Doublon momentum distribution function for times $t=0,1,2,3,4,5,10,20\, \tau_{dh}$ after the quench, where $\tau_{dh}$ is the doublon-holon scattering time, and $ \rho = 4 (J/U)^2 $ is the doublon density.
 For $t \to \infty$ an equilibrium state at $T=\infty$ is approached where all momenta are equally occupied.
\label{fig2} }
 \end{figure}
To describe relaxation to equilibrium after the quench, we first calculate the probability,  $p_k=8 \, (J/U)^2 \sin^2 k=\rho \, (1- \cos(2k))$, that
a doublon-holon pair with  momenta $k$ and $-k$ is created at a given site. This probability is used to create
at $t=0$  an ensemble of doublons and holons moving with the velocity $\partial_k \epsilon^d_k$ and $\partial_k \epsilon^h_k$, respectively. This allows to determine position and time of the next scattering event.
Doublon-doublon and holon-holon scattering has no effect (as only the momenta of the two particles are exchanged), while the scattering of holons and doublons leads to relaxation. To leading order in $U\gg J$
 the reflection probability is $1$ and the new momenta after scattering can just be calculated from energy and lattice-momentum conservation. Repeating this procedure, we track the motion of about $10^5$ quasiparticles
for long times and, furthermore, average over $ 500 $ ensembles. Correlation functions obtained by this semiclassical dynamics are expected to give the corresponding quantum mechanical correlation functions for $U \gg J$ \cite{Sachdev97}.
Related models of hard-core particles moving in $1d$ have also been simulated, e.g., in Ref. \cite{Grassb02,Cipriani05,Delfini07}
to study thermal transport. In contrast to our case, however, quadratic dispersions and momentum conservation was
used. We have also implemented a version which takes into account the finite tunneling probability of doublons and holons of order $(J/U)^2$, but
as qualitatively similar results have been obtained in this case, we only show results for vanishing tunneling rate
in the following.

In Fig.~\ref{fig2} we show how the momentum distribution of doublons gets flatter and flatter as a function
of time. As the initial total kinetic energy is on average zero, the system relaxes towards an equilibrium state at $T=\infty$. For all semiclassical plots we measure the timescale in units of
$\tau_{dh}\approx 0.031 \, U^2/J^3 = 0.123 \rho^{-1}/J$, the doublon-holon scattering time (obtained by dividing the simulated time by the total number of doublon-holon scattering events and the number of doublons). In these units, all semiclassical results are completely independent of $U/J$.

Note that the semiclassical approach does not contain extremely rare
processes where
a holon/doublon pair is created or annihilated by converting the kinetic energy of a large number of quasiparticles (of order $U/J$)  into interaction energy  in a complicated process~\cite{doublon10}. We will not consider the exponentially (in $U/J$) long time scales  \cite{doublon10}  where these processes become important and
which ultimately lead to an equilibrium state with  $J \ll T<\infty$.

\begin{figure}[t]
\includegraphics[width=0.9 \linewidth]{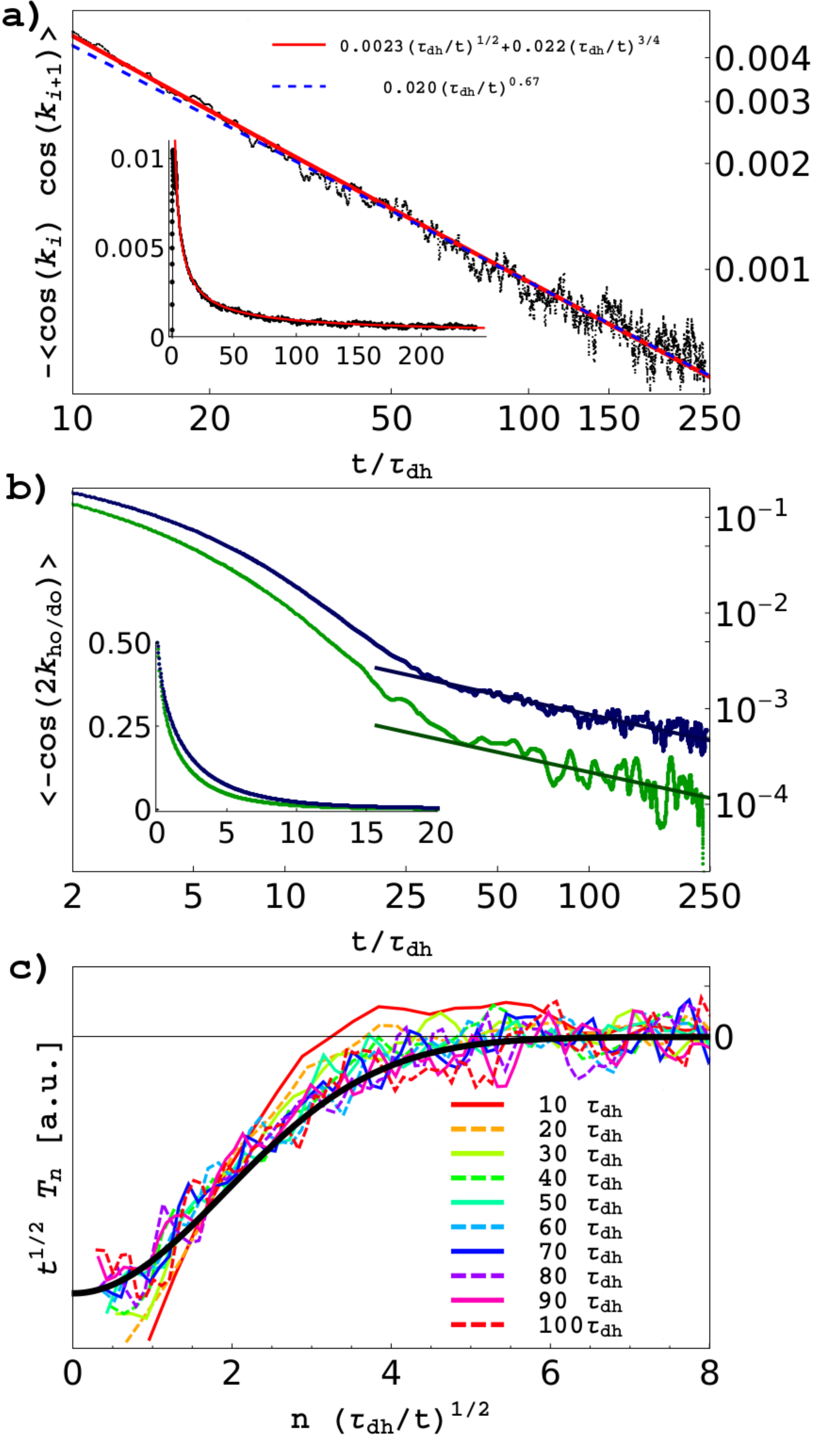}
\caption{a) The decay of the energy-energy correlation function of nearest neighbor quasiparticles is very slow in time and consistent with hydrodynamic predictions.
b) Expectation value of $- \cos (2 k)$ for doublons (blue) and holons (green). Initially these quantities decay approximately exponentially in a few collisions. At long times, however, a hydrodynamic long-time tail is visible.
The solid lines are quantitative predictions of these long-time tails, obtained from the fit to the upper plot using Eq. (\ref{pre}).
c) Scaling plot of $T_n$, Eq.~(\ref{Tnn}). The thick line is Eq.~\ref{fita} with $D_e$ evaluated from the Kubo formula.\label{fig3}}
\end{figure}

While we will argue that hydrodynamic long time tails generically govern the relaxation of most physical observables for $t\to \infty$, we find that they are much more pronounced in some observables.
We obtain the  most pronounced long-time times for the correlation functions
\begin{align} \label{Tnn}
T_n(t) = \langle \cos k_i(t)  \cos k_{i+n}(t) \rangle
\end{align}
describing the energy-energy correlation function of particle $i$ and particle $i+n$ (we numerate the particles from left to right).
Within a simple Boltzmann equation treatment of the problem (see supplement \cite{supplement}), this quantity vanishes.
The solid line in Fig.~\ref{fig3}a shows a fit to $ T_1(t) $ of the form $c_1/t^{1/2}+c_2/t^{3/4}$
consistent with Eq.~(\ref{tails}). Note that it is mandatory to include subleading corrections to the fit
as those are only suppressed by relative factors of $1/t^{1/4}$.
From a simple power-law fit, $c t^{-\alpha}$ (dashed line), one obtains $\alpha \approx 0.67$ which describes the numerical data equally well. While this exponent is reminiscent of the $2/3$ expected for the KPZ universality class, we believe that this agreement is only accidental as
strong Umklapp scattering relaxes the current rapidly inconsistent with the KPZ universality class, see above.
At least three orders of magnitude longer simulations are needed to be able to distinguish numerically the different asymptotic behavior.
A similar discussion of an equilibrium correlation function is given in the supplement~\cite{supplement}.

To investigate, how the hydrodynamic correlations spread in space, we show
in Fig.~\ref{fig3}c a scaling plot of $\sqrt{t} T_n(t)$ as a function of $n/\sqrt{t}$
 at different times. The approximate scaling collapse for long times shows that the information
spreads diffusively, $n \sim \sqrt{t}$. From linear hydrodynamics, Eq. (\ref{diffusion}), one can easily calculate
the scaling function \cite{supplement}
\begin{equation}
 T_n(t) \sim \frac{1}{\sqrt{t}} \exp ( -\frac{n^2}{8 \tilde{D}_e  t/\tau_{dh}})\label{fita}
\end{equation}
While the prefactor, depending on details of the quench, is unknown, the energy diffusion constant $\tilde D_e=D_e \tau_{dh} \rho^2\approx 0.91$ can be calculated
from the Kubo formula evaluated at thermal equilibrium. The thick black line in Fig.~\ref{fig3}c
shows that the analytic formula describes the data quantitatively. This shows that indeed linear hydrodynamics
governs the buildup of long-time tails.

When investigating the relaxation of the momentum distribution, such long-time tails are much more difficult to detect. In  Fig.~\ref{fig3}b, we show the expectation value of $\cos(2 k)$ for doublons and holons
\begin{align}
T^{d/h}(t)= \langle \cos(2 k^{d/h}_i)\rangle
\end{align}
where the expectation value is determined by summing only over doublon or holon momenta, respectively. Up to a normalizing factor, this is the dominant Fourier component of the
distribution function shown in Fig.~\ref{fig3}b. As $\cos^2 k=(1+\cos 2k)/2$, it is also directly related
to the square of the kinetic energy of each particle. At first, $T^{d/h}$ decays exponentially on time scales consistent with predictions from the Boltzmann equation \cite{supplement}. For $t \gtrsim 25\,\tau_{dh}$, however, a small but finite long-time tail dominates the relaxation.

The relative prefactors of the long-time tail in various physical quantities are related to each other as they arise from the same hydrodynamic modes. To predict analytically how the prefactors are related, we recall their physical origin: after a {\em local} equilibrium has been established, it takes a long time to establish {\em globally} the characteristic fluctuations of the conserved densities. We therefore investigate first how {\em in thermal equilibrium}
the observables $T_{1,{\rm eq}}, T_{\rm eq}^d$ and $T_{\rm eq}^h$ depend on the densities $n^d$, $n^h$ and the energy per particle $\epsilon$, using that in the semiclassical limit the distribution functions are given
by $n_k^{d/h}=z_{d/h} e^{-\beta \epsilon_k^{d/h}}$, where the fugacities $z_{d}$, $z_h$ and $\beta=1/T$ are functions of $n^d$, $n^h$ and $\epsilon$ .
In the considered semiclassical limit $T_{1,{\rm eq}}, T_{\rm eq}^d$ and $T_{\rm eq}^h$ are independent of $n^d$ and $n^h$, so that we obtain in equilibrium to leading order in $ \epsilon$
\begin{eqnarray} \label{eqlbr}
T_{1,{\rm eq}}&\approx& \frac{17}{200} \left(\frac{\epsilon}{J}\right)^2 \nonumber \\
 T_{\rm eq}^d &\approx& \frac{16}{200} \left(\frac{\epsilon}{J}\right)^2, \qquad  T_{\rm eq}^h \approx \frac{4}{200} \left(\frac{\epsilon}{J}\right)^2
\end{eqnarray}
As $\epsilon=0$ on average, the above determined prefactors describe how sensitive the observables are
to fluctuations of the energy. This allows to predict that
 \begin{eqnarray}\label{pre}
T^d=\frac{16}{17} \, T_1, \qquad T^h=\frac{4}{17} T_1 \qquad \text{for } t\to\infty
\end{eqnarray}
fully consistent with our numerical results, as shown by the solid lines in Fig.~\ref{fig3}.

\begin{figure}[t]
\includegraphics[width=0.9 \linewidth]{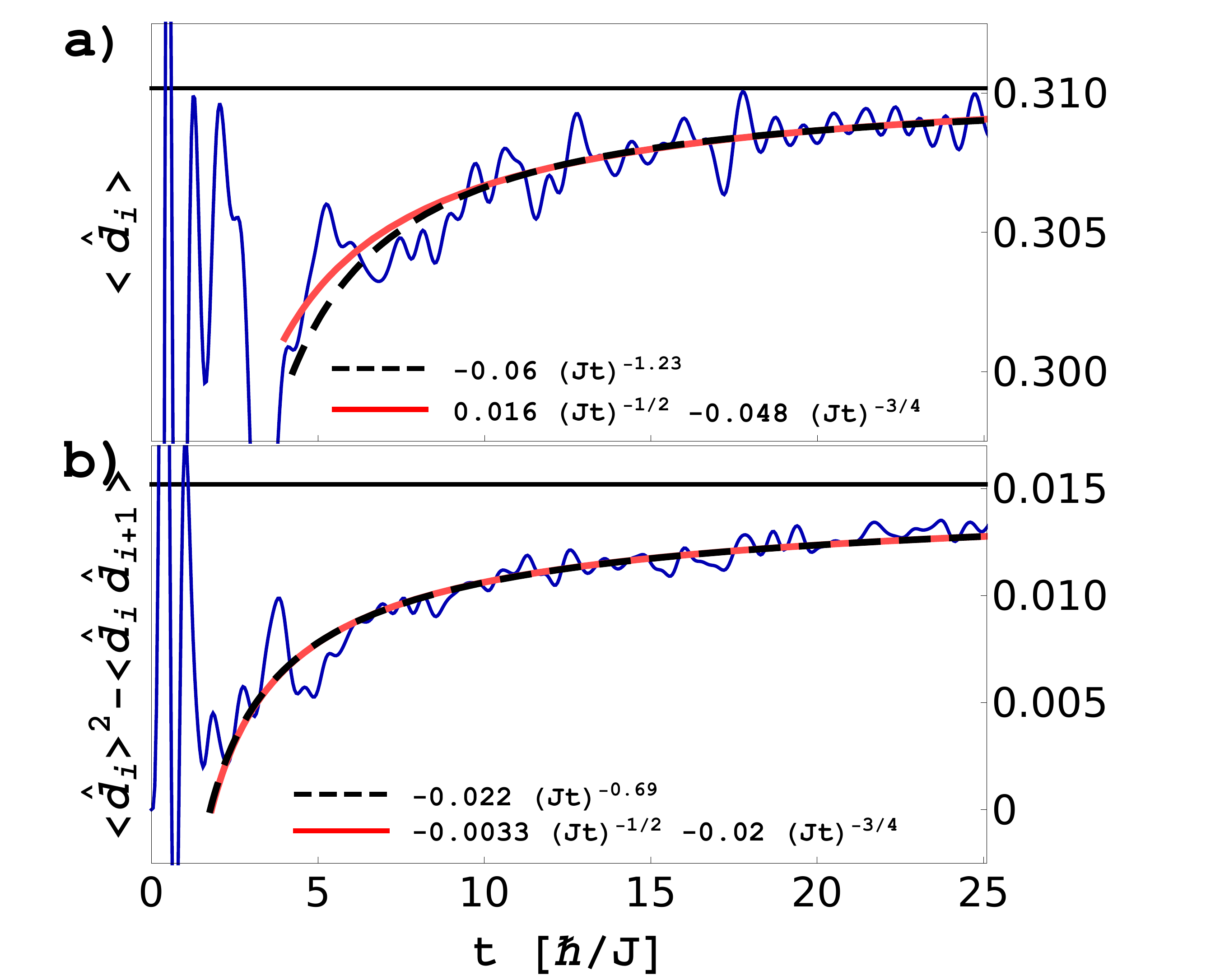}
\caption{Time dependence of the density of doubly occupied sites , $\langle \hat d_i \rangle= \langle n_i (n_i-1)\rangle/2$, and of their nearest-neighbor correlation function, $-\langle \hat d_i \hat d_{i+1}\rangle+\langle \hat d_i \rangle^2$ obtained
from exact diagonalization of a $14$ site system after a quench from $U=\infty$ to $U=J$. The straight solid line denotes the infinite-time average (obtained from the diagonal ensemble).
 \label{fig5} }
\end{figure}
The semiclassical approach discussed above breaks down for strong quenches when $U\sim J$. In this regime
we have used the ALPS code \cite{alps} for brute-force exact diagonalization of small systems (excluding bosonic occupations $\ge 3$ \cite{supplement}) to investigate the relaxation to equilibrium.
Even for this fully quantum mechanical calculation we obtain strong evidence of hydrodynamic long-time tails, see
Fig.~\ref{fig5}. The
clearest signature is obtained for the nearest-neighbor correlation function of doubly occupied sites, Fig.~\ref{fig5}b,
which is one contribution to the energy-energy correlation function. But also the number of doubly occupied sites (Fig.~\ref{fig5}a), which can directly be measured experimentally \cite{doublon10},
shows pronounced long-time tails. From the numerical results for such small systems it is impossible to extract
any exponents as shown by the various fits in Fig.~\ref{fig5}.

It is surprising that the slow relaxation is so pronounced already in 14-sites systems: often finite size effects dominate for such small systems and long times.
Here it helps that diffusive transport is much slower than ballistic transport. The typical time scale associated with diffusive transport over $N$ sites scales with $N^{2}$. As $14^{2}=196$, it is not surprising that diffusive effects dominate on the simulated time scales, see supplement \cite{supplement} for a study of the system size dependence.

Global equilibration characterized by slow power-law relaxation with characteristic long-time tails is of importance
for practically all physical observables in generic quantum and classical many-particle systems as long as a
small number of conservation laws are present. Our results have, however, shown that the experimental importance of these long-time tails depend strongly on the observable which is considered.
Sometimes they are difficult to observe due to tiny prefactors. Unexpectedly,
the long-time tails turned out to be more pronounced in the quantum regime compared to the
semi-classical limit, possibly due to the more complex interaction processes in a many-particle quantum system.

Controlling global equilibration is a key element for present and future high-precision experiments with ultra-cold atoms. As loss processes often put severe limits on achievable time scales, it is important to design experimental sequences in such a way that the desired state of matter can be reached rapidly. Speeding up global equilibration, e.g.,  by partially suppressing scattering and therefore enhancing transport, can be an important element of such a strategy.

{\sl Acknowledgements:} It is a pleasure to thank M. Becker, P. Br\"ocker, J. Krug, W. Michel and, especially, H. Spohn for helpful discussions. This work was supported by  NSF-DMR 1303177 (AM), the
Simons Foundation (AM), and the SFB TR12 of the DFG.

\bibliography{bib}
\newpage

\renewcommand{\theequation}{S\arabic{equation}}
\renewcommand{\thefigure}{S\arabic{figure}}

\setcounter{equation}{0}
\setcounter{figure}{0}
\onecolumngrid

{\Large \center {\bf Supplementary Material to: Hydrodynamic long-time tails after a quantum quench}} \\

In this supplement, we review how long time tails arise in linear hydrodynamics, we
provide further information on the numerical implementation and the finite-size dependence of the
exact diagonalization. Furthermore, we investigate to what extent the Boltzmann equation describes the semiclassical regime and
discuss long time tails of an equilibrium correlation function.

\section{Long-time tails in linear hydrodynamics}

In this section we briefly review the origin of long-time tails after a quench in systems without momentum conservation.
While all results presented in this section are well known, we have not been able to find an appropriate reference.
For simplicity we restrict the discussion to a single conserved quantity, the energy.
The hydrodynamic equations can easily be generalized to several diffusive modes by replacing the energy density $e$ by a vector of conserved densities,
the diffusion constant by a diffusion tensor and the specific heat by a matrix of thermodynamics susceptibilities  (For the semiclassical calculation
 discussed in the main text energy does not couple to the other conservation laws, see below).
The qualitative results remain unmodified as long as all diffusion constants are finite, only conservation laws even under
time reversal and parity are considered and, most importantly, momentum is not conserved.

The starting point is the linearized stochastic diffusion equation in $d$ dimensions
\begin{equation}\label{linD}
\partial_t e - D_e \mathbb{\nabla}^2 e=\mathbb{\nabla} \bf f
\end{equation}
where $D_e$ is the energy diffusion constant and $f_i$ describes thermal fluctuations of the energy current.
The size of fluctuations can be determined from the condition that the equilibrium correlations of the energy are correctly reproduced by Eq. (\ref{linD})
\begin{equation}\label{equi}
\langle e(r) e(r') \rangle_{\rm eq}=c_V k_B T^2 \delta(r-r')
\end{equation}
where $c_V$ is the the specific heat per volume.
In reality, these correlations are not exactly local but as the hydrodynamic equations describe only the behavior at long time and length scales, we can approximate the spatial correlations by a $\delta$ function.
One obtains therefore that the fluctuations of the current
\begin{equation}
\langle f_i(r,t) f_j(r',t')\rangle=\delta_{ij} 2 k_B T^2 c_V D_e \delta(r-r') \delta(t-t')\label{corr}
\end{equation}
are proportional to both the diffusion constant and the specific heat.

For a given initial condition, $e(r,0)=e_0(r)$, Eq. (\ref{linD}) is solved for $ t \geq 0 $ by
\begin{eqnarray}
e(r,t)&=&\int  d^dr' \, g_{D_e}(r-r',t) e_0(r')  + \int_0^t dt' \int d^dr'\, g_{D_e}(r-r',t-t') \mathbb{\nabla} {\bf f}(r',t') \\
g_{D_e}(r,t)&=&\frac{1}{(4 \pi {D_e} t)^{d/2}} \exp\!\left(-\frac{r^2}{4 D_e t}\right)\label{diffG}
\end{eqnarray}
From this solution and Eq. (\ref{corr}) one obtains directly the relaxation of the energy fluctuations as a function of time
\begin{eqnarray}\label{diffeql}
\langle e(r,t) e(r',t) \rangle -\langle e(r) e(r') \rangle_{\rm eq} &=&\int  d^d r_1 d^d r_2 \  g_{D_e}(r-r_1,t)g_{D_e}(r'-r_2,t) \left(\langle e_0(r_1) e_0(r_2) \rangle -\langle e(r_1) e(r_2) \rangle_{\rm eq}\right)
\end{eqnarray}
This equation describes how the fluctuations of energy approach their equilibrium value. As generically the energy fluctuations directly after a quench, $\langle e_0(r) e_0(r') \rangle$, will differ from their expectation value in the long time limit, they have to be built up slowly by diffusive transport.

Assuming sufficiently short-ranged correlations in the initial state and using that $\int dr_1  g_D(r-r_1,t)g_D(r'-r_1,t) = g_{2D}(r-r',t)$ , one obtains from Eqs.~(\ref{diffG}) and (\ref{diffeql})
\begin{equation} \label{eneq}
\langle e(r,t) e(r',t) \rangle -\langle e(r) e(r') \rangle_{\rm eq} \sim \frac{1}{t^{d/2}} \exp \left( -\frac{(r-r')^2}{8 D_e t} \right)
\end{equation}
After a quench, one approaches therefore the global equilibrium state only algebraically. Corrections to this formula from non-linear contributions are discussed below.
In Fig.~3 of the main text, we show that the spread of correlations in our semiclassical simulations indeed follows Eq. (\ref{eneq}), see solid line in Fig. 3c.
To obtain a quantitative fit, we have determined the diffusion constant $ D_e $ from heat conductivity $\kappa$ using
\begin{equation}
 D_e = \kappa \left( \frac{\partial \langle e \rangle_{\rm eq}}{\partial T} \right)^{-1}=\kappa \left( \frac{1}{k_B T^2} \sum_i \rho_i \langle e_i^2 \rangle_{\rm eq}\right)^{-1}
\end{equation}
where the sum runs over the two particle species with density $\rho_i$.
The thermal conductivity $\kappa$ can be determined numerically using the Kubo formula {[}S1{]}
\begin{equation}\label{heatc}
 \kappa = \frac{1}{L k_B T^2} \int\limits_0^\infty dt \langle J_e(t) J_e(0)\rangle_{\rm eq}
\end{equation}
where $ J_e $ is the total energy current, and $L$ is the system size. For our semiclassical simulations we have to consider the $T\to \infty$ limit. Note that $D_e$ is finite in this limit
as all factors of $T$ cancel. Furthermore, all linear thermoelectric effects, i.e., the coupling of the energy current to gradients of the particle density vanish 
in this limit (non-linear couplings do, however, exist, see below)
which justifies the use energy diffusion only in the derivation of Eq.~(\ref{eneq}).

To estimate the importance of corrections to the linear stochastic diffusion equation,
it is useful to perform a simple scaling analysis. Eq.~(\ref{linD}) is invariant under the scaling transformation
$x\to \tilde x, t\to \tilde t, f\to \tilde f$ and $e \to \tilde e$ with
\begin{equation}
x=\lambda \tilde x,\quad t=\lambda^2 \tilde t, \quad e=\frac{1}{\lambda^{d/2}} \tilde e, \quad f= \frac{1}{\lambda^{(d+2)/2}} \tilde f
\end{equation}
The analog scaling relations also apply for the density $n$ and the fluctuations of the charge current.
Examples of possible correction terms are $\alpha \nabla( e \nabla e)$, $\alpha' \nabla( n \nabla e)$ or
$\beta \nabla^4 e$. Rewriting those in the new variables, one finds that they are suppressed for large $\lambda$,
$\tilde \alpha = \alpha/\lambda^{d/2}$, $\tilde \alpha' = \alpha/\lambda^{d/2}$ and $\tilde \beta = \beta/\lambda^2$ (reflecting the scaling dimensions
of $e$ and $\nabla^2$, respectively). The relaxation of after a quantum quench is therefore expected to be of the form
\begin{equation}
 \langle e(r,t) e(r',t) \rangle -\langle e(r) e(r') \rangle_{\rm eq} \sim \frac{1}{\sqrt{t}} f\!\left(\frac{r}{\lambda},\frac{t}{\lambda^2},
\frac{\alpha}{\lambda^{d/2}}, \frac{\alpha'}{\lambda^{d/2}}, \frac{\beta}{\lambda^2}, \dots \right)=\frac{1}{\sqrt{t}} f\!\left(\frac{r}{\sqrt{t}},1,
\frac{\alpha}{t^{d/4}}, \frac{\alpha'}{t^{d/4}}, \frac{\beta}{t},\dots \right)
\end{equation}
where $f$ is a scaling function, the dots denote further subleading corrections, and we have set $\lambda=\sqrt{t}$ in the last equality. Using a Taylor expansion in the last three arguments for large $t$,
one  finds that corrections  are suppressed 
by $\alpha/t^{d/4}$, $\alpha'/t^{d/4}$ and $\beta/t$. In our model, the $\alpha$ term is absent at $T=\infty$ due to a $e \to -e$ symmetry. The $\alpha'$ term, 
however, should be present. Note that non-linearities of the form
$\nabla e^2$ arising as corrections to the momentum current in systems with momentum conservation are
instead relevant perturbations in $d=1$ leading to the KPZ universality class {[}S2,S3{]}. Such a term cannot arise as a correction to 
the energy or charge current as it would violate inversion symmetry. A related discussion of how subleading non-linearities affect the optical conductivity of
metals without momentum conservation has been given in Ref.~S4.

\begin{figure}[t]
\includegraphics[width=0.7 \linewidth]{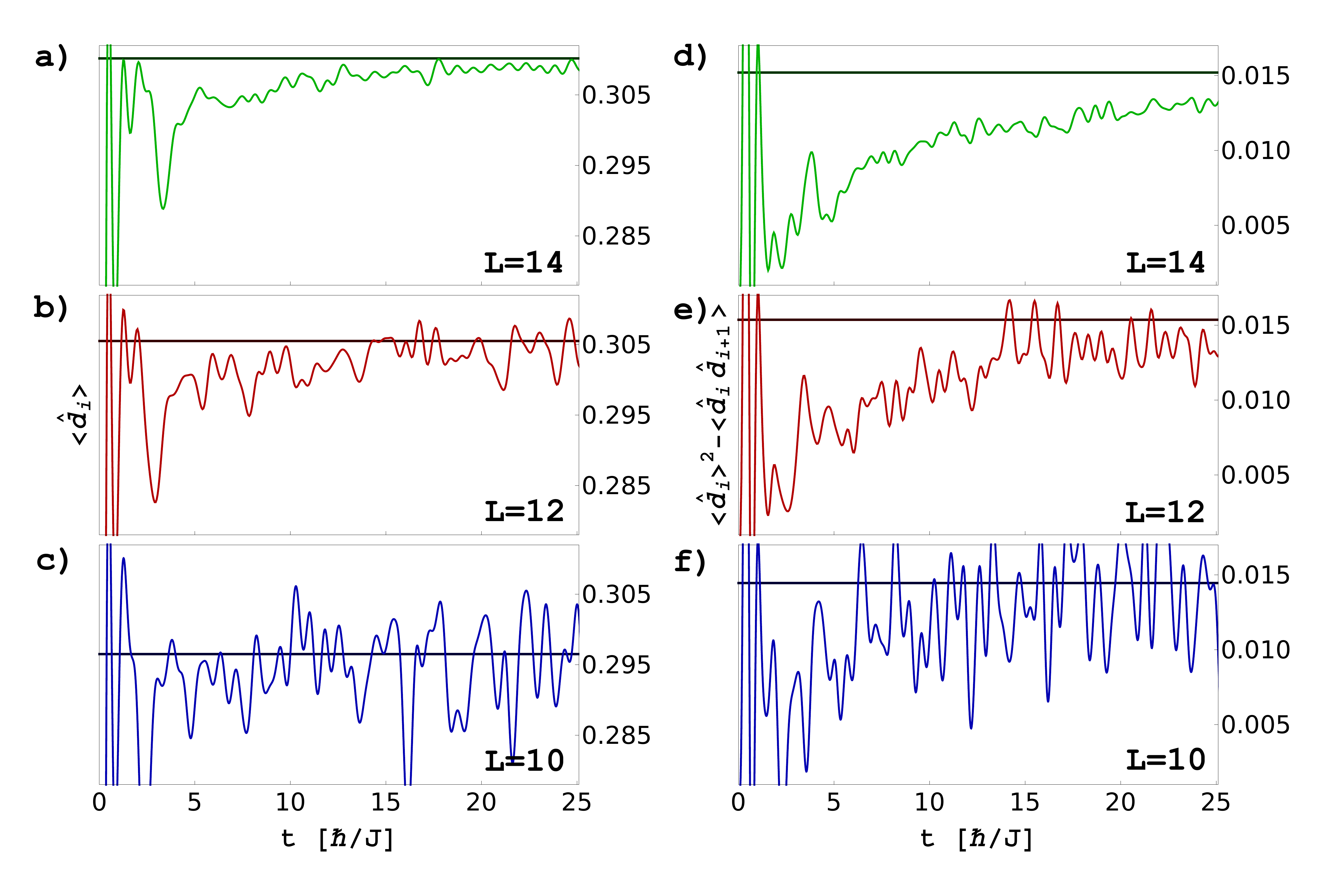}
\caption{Exact diagonalization data for system sizes $ L=14,12,10 $ after a quench from $U=\infty$ to $U=1$. The solid lines are the long-time averages obtained for each system size separately.
a-c) Number of doublons per site,   $\langle \hat d_i \rangle= \langle n_i (n_i-1)\rangle/2$.
d-f) Nearest neighbor doublon correlations,  $-\langle \hat d_i \hat d_{i+1}\rangle+\langle \hat d_i \rangle^2$ with $\hat d_i =   n_i (n_i-1)/2$.
 \label{figSED} }
\end{figure}

\section{Finite size effects in exact diagonalization}
The exact diagonalization data for the bosonic Hubbard model, Eq.~(4) of the main text, were obtained for $ U = J $ using the ALPS code {[}S5{]}.
The initial state, a perfect Mott insulator ($ U = \infty $ or $ J =0 $), is a product state where each site is singly occupied.
This corresponds to a strong quench -- the initial state is not close to an eigenstate.
We use periodic boundary conditions, and we do not include states in the Hilbert space where $3$ or more bosons occupy a single site.
Therefore our results describe a modified Hubbard model with a large 3-particle term $U' \sum_i n_i (n_i -1) (n_i-2)$.
This allows one to reach larger system sizes.

In Fig.~\ref{figSED} the observables of Fig.~4 from the main text are shown for different system sizes $ N=10,12,14 $.
While for $ N=10$ the finite size fluctuations do not allow one to see how the long time average is approached, the power law behavior
can clearly be seen for $ N =14 $. The system sizes are, however, too small for a meaningful finite size scaling analysis.

\section{Boltzmann equation in the semiclassical regime}

For weak quenches, $ U \gg J $, the density of of excitations, doublons and holons, induced by the quench is very low. It is instructive to investigate their relaxation within the Boltzmann approach.
This approach can describe local equilibration but does not  reproduce hydrodynamic long-time tails in homogeneous systems.

Denoting the semiclassical distribution functions of doublons and holons as $n_k^d$ and $n_k^h$, respectively, the Boltzmann equation takes the form
\begin{eqnarray}
\frac{\partial}{\partial t} n^d_k &=&\int \frac{d q}{2 \pi} \int \frac{d k'}{2 \pi} \int \frac{d q'}{2 \pi}W_{k,q;k',q'} \delta\!\left( \epsilon^d_k+\epsilon^h_q - (\epsilon^d_{k'}+\epsilon^h_{q'}) \right) \delta_U\! \left( k+q-(k'+q') \right)
\left( n^d_{k'} n^h_{q'}- n^d_k n^h_{q} \right) \label{bmeqd}\\
\frac{\partial}{\partial t} n^h_q &=&\int \frac{d k}{2 \pi} \int \frac{d k'}{2 \pi} \int \frac{d q'}{2 \pi}W_{k,q;k',q'} \delta\!\left( \epsilon^d_k+\epsilon^h_q - (\epsilon^d_{k'}+\epsilon^h_{q'}) \right) \delta_U\! \left( k+q-(k'+q') \right)
\left( n^d_{k'} n^h_{q'}- n^d_k n^h_{q} \right) \label{bmeqh}
\end{eqnarray}
where $\delta_U(k)=\sum_n \delta(k+ n 2 \pi)$ as Umklapp scattering can relax the momentum by multiples of the reciprocal lattice vector.
The transition rate for hard-core collisions in 1 dimension is exactly given by
\begin{eqnarray}
W_{k,q;k',q'} &=& (2 \pi)^2 |\partial_k \epsilon^d_k - \partial_q \epsilon^h_q| |\partial_{k'} \epsilon^d_{k'} - \partial_{q'} \epsilon^h_{q'}|
\end{eqnarray}
After using energy and lattice momentum conservation and performing the Fourier series expansion of the distribution functions as
\begin{eqnarray}
 n^{d/h}_k = \sum\limits_m \cos(m k) \; d_m / h_m
\end{eqnarray}
we find equations for the Fourier components
\begin{eqnarray}
\frac{\partial}{\partial t} d_m & = & 2 \sum\limits_{m',m''} d_{m'} h_{m''} \, \int \frac{d k}{2 \pi} \int \frac{d q}{2 \pi} \cos(m k) |\partial_k \epsilon^d_k - \partial_q \epsilon^h_q| \left( \cos(m' k_d) \cos(m'' (k+q- k_d)) -\cos(m' k) \cos(m'' q) \right) \nonumber \\
\frac{\partial}{\partial t} h_m & = & 2 \sum\limits_{m',m''} d_{m'} h_{m''} \, \int \frac{d k}{2 \pi} \int \frac{d q}{2 \pi} \cos(m q) |\partial_k \epsilon^d_k - \partial_q \epsilon^h_q| \left( \cos(m' k_d) \cos(m'' (k+q- k_d)) -\cos(m' k) \cos(m'' q) \right) \nonumber
\end{eqnarray}
where $ k_d (k,q) $ is the doublon momentum after the scattering of a doublon with momentum $ k $ and a holon with momentum $ q $, determined from energy and
momentum conservation (modulo Umklapp scattering).

\begin{figure}[t]
\includegraphics[width=0.7 \linewidth]{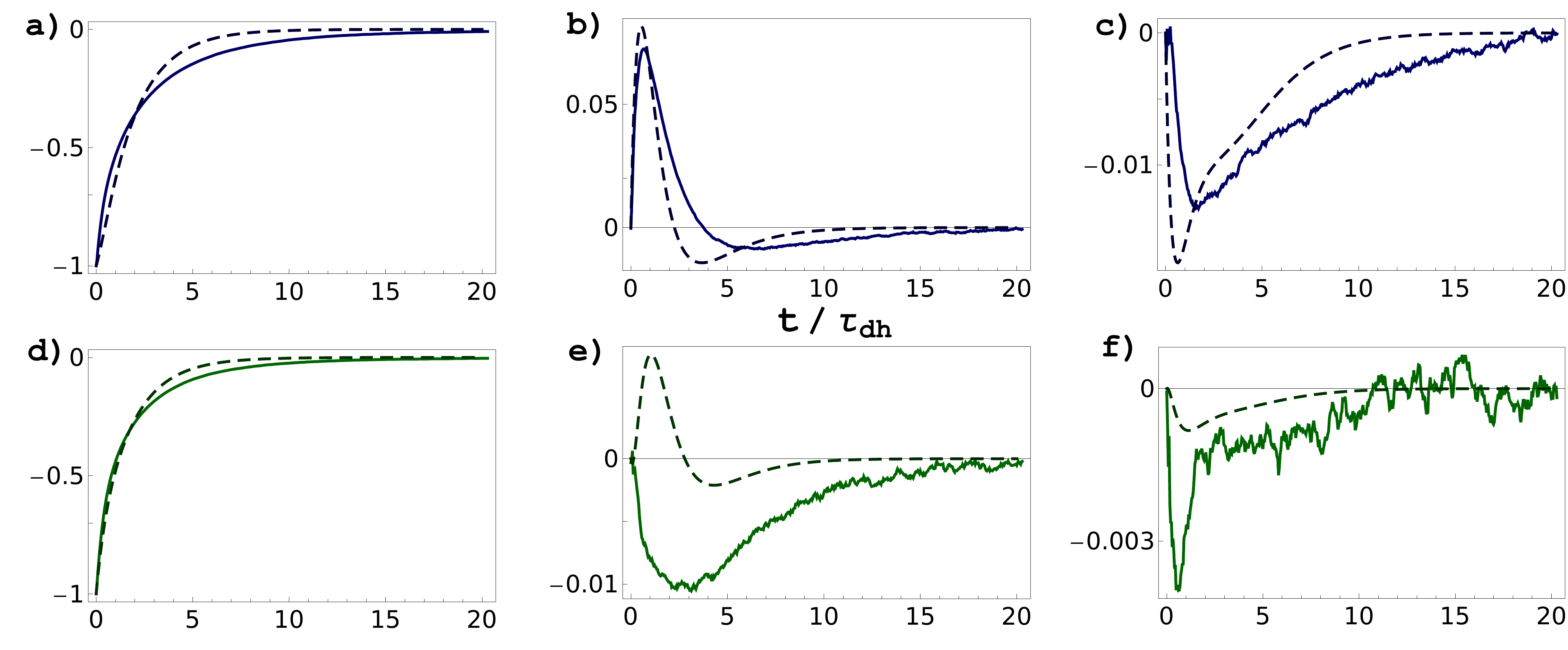}
\caption{Relaxation of $\langle \cos( n k)\rangle$, $n=2,4,6$ for doublons (plot a-c) and holons (plot d-f).
Solid line: simulation data, dashed line: Boltzmann equation result.
 \label{figSbmeq} }
\end{figure}

We have solved Eqs.~(\ref{bmeqd}) and (\ref{bmeqh}) numerically, including all modes up to $ m=8 $, with the initial conditions $ h_0 (0) = d_0 (0) = - h_2 (0) = - d_2 (0) = \rho $ and $ h_m(0) = d_m(0) =0 $ otherwise, to match the initial condition
$n_k^d (0)=n_k^h (0) = 8 \, (J/U)^2 \sin^2 k $, see Fig.~2 of the main text.
The result for the even modes (the odd modes are all zero) is compared in Fig.~\ref{figSbmeq} to the simulation data.
The Boltzmann equation predicts correctly the times scale of relaxation, see Figs.~\ref{figSbmeq}a and \ref{figSbmeq}d.
As in $d=1$ there is a high probability that one particle scatters again and again with the same scattering partner, a full quantitative agreement cannot be expected even for the first few scattering events.

For long times the Boltzmann equation predicts exponential relaxation as it does not capture thermal fluctuations of energy and particle density but describes those only on average.
\section{Equilibrium correlation function}

\begin{figure}[h!]
\includegraphics[width=0.5 \linewidth]{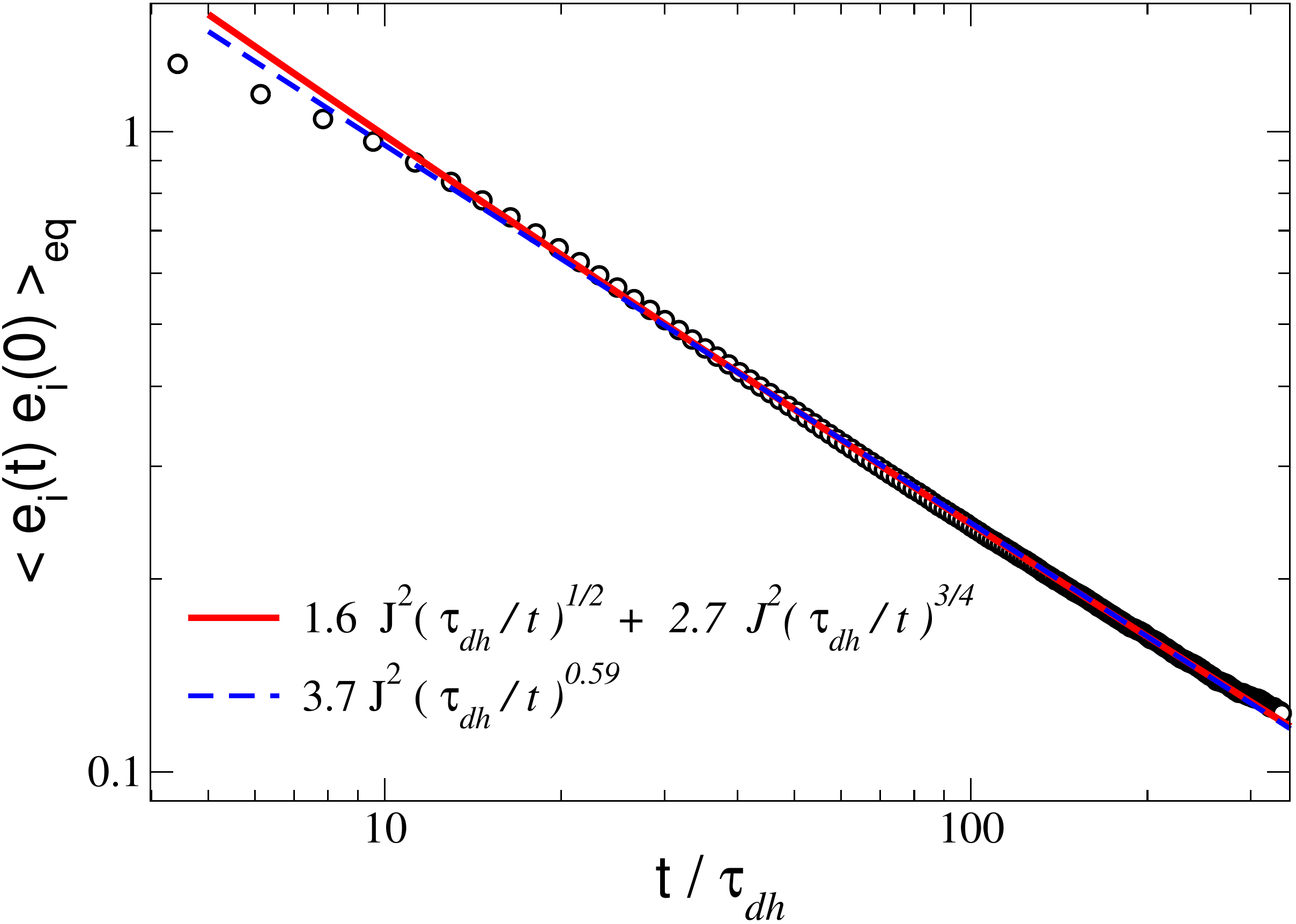}
\caption{\label{figSeql} Correlation function $\langle e_i(t) e_i(0)\rangle_{\rm eq}$ calculated in thermal equilibrium together with two fits to this function, see legend, which equally well describe the numerical result. }
\end{figure}

Hydrodynamic long-time tails also dominate equilibrium correlation functions. In Fig.~\ref{figSeql} we show
$\langle e_i(t) e_i(0)\rangle_{eq}$ obtained from a semiclassical simulation in equilibrium at $T=\infty$ where initially
all doublons and holons have a $k$-independent momentum distribution $n^d_k=n^h_k={\rm constant}$ and are uncorrelated in space. We assume the same density of doublons and holons. As for the quenches studied in the main text, it is difficult to extract reliably the long-time asymptotics due to large subleading correction which vanish only slowly.
The numerical data is equally well described by a power-law fit with exponent $0.59$ and a fit to $c_1 t^{-1/2} + c_2 t^{-3/4}$ expected
from hydrodynamics.
\newline
{[}S1{]}  R.~Kubo, M.~Yokota, and S.~Nakajima, J.~of Phys.~Soc.~of Japan {\bf 12}, 1203 (1957). \\
{[}S2{]}  M.~Kardar, G.~Parisi, and Y.-C.~Zhang, Phys.~Rev.~Lett. {\bf 56}, 889 (1986). \\
{[}S3{]}  H.~Spohn, arXiv:1305.6412 (2013). \\
{[}S4{]}  S.~Mukerjee, V.~Oganesyan, and D.~Huse, Phys.~Rev.~B {\bf 73}, 035113 (2006).\\
{[}S5{]}  B.~Bauer, {\it et al.} J.~of Stat.~Mech.: Theory and Exp. {\bf 2011} (2011).

\end{document}